\documentclass[%
reprint,
 amsmath,amssymb,
 aps,
]{revtex4-1}

\usepackage{graphicx}
\usepackage{dcolumn}
\usepackage{bm}

\begin{document}

\preprint{APS/123-QED}

\title{$^{77}$Se NMR Investigation of the K$_{x}$Fe$_{2-y}$Se$_{2}$ High $T_{c}$ Superconductor ($T_{c}=33$~K)}
\thanks{The first two authors made equal contributions to this work.}%

\author{D. A. Torchetti,$^{1}$ M. Fu,$^{1}$ D. C. Christensen,$^{1}$ K. J. Nelson,$^{1}$ T. Imai,$^{1,2}$ H.\ C.\ Lei$^{3}$ and C. Petrovic$^{2,3}$ }
\affiliation{%
$^{1}$Department of Physics and Astronomy, McMaster University, Hamilton, Ontario L8S 4M1, Canada
}%
\affiliation{%
$^{2}$Canadian Institute for Advanced Research, Toronto M5G 1Z8, Canada
}%
\affiliation{%
$^{3}$Condensed Matter Physics and Materials Science Department, Brookhaven National Laboratory, Upton, New York 11973, USA
}%

\date{\today}

\begin{abstract}
We report comprehensive $^{77}$Se NMR measurements on a single crystalline sample of the newly discovered FeSe-based high temperature superconductor K$_{x}$Fe$_{2-y}$Se$_{2}$ ($T_{c}=33$\ K) in a broad temperature range up to 290~K.  Despite deviations from the stoichiometric KFe$_{2}$Se$_{2}$ composition, we observed $^{77}$Se NMR lineshapes as narrow as 4.5~kHz under a magnetic field applied along the crystal c-axis, and found no evidence for co-existence of magnetic order with superconductivity.  On the other hand, the  $^{77}$Se NMR lineshape splits into two peaks with equal intensities at all temperatures when we apply the magnetic field along the ab-plane.  This suggests that K vacancies may have a superstructure, and that the local symmetry of the Se sites is lower than the tetragonal four-fold symmetry of the average structure.  This effect might be a prerequisite for stabilizing the s$_{\pm}$ symmetry of superconductivity in the absence of the hole bands at the Brillouin zone center.  From the increase of NMR linewidth below $T_c$ induced by the Abrikosov lattice of superconducting vortices, we estimate the in-plane penetration depth $\lambda_{ab} \sim 290$~nm, and the carrier concentration $n_{e} \sim 1 \times 10^{+21}$~cm$^{-3}$.  Our Knight shift $^{77}K$ data indicate that the uniform spin susceptibility decreases progressively with temperature, in analogy with the case of FeSe ($T_{c}\sim9$\ K) as well as other FeAs high $T_{c}$ systems.  The strong suppression of $^{77}K$ observed immediately below $T_c$ for all crystal orientations is consistent with a singlet pairing of Cooper pairs.  We don't, however, observe the Hebel-Slichter coherence peak of the nuclear spin-lattice relaxation rate $1/T_{1}$ immediately below $T_{c}$, expected for conventional BCS s-wave superconductors.  In contrast with the case of FeSe, we do not observe evidence for an enhancement of low frequency antiferromagnetic spin fluctuations (AFSF) near $T_c$ in $1/T_{1}T$.  Instead, $1/T_{1}T$ exhibits qualitatively the same behavior as overdoped non-superconducting Ba(Fe$_{1-x}$Co$_{x}$)$_{2}$As$_{2}$ with $x \sim 0.14$ or greater, where hole bands are missing in the Brillouin zone center.  We will discuss the implications of our results on the hitherto unknown mechanism of high temperature superconductivity in FeSe and FeAs systems.
\end{abstract}

\maketitle


\section{\label{sec:level1}Introduction}
The mechanism of high temperature superconductivity in iron-arsenides \cite{Kamihara} remains highly controversial despite a large volume of experimental and theoretical works reported over the past three years \cite{Norman, PhysicsToday, Johnston, Greene}.  More thorough and systematic investigations of various iron-based systems are necessary to identify the superconducting mechanism.  Very recently, Guo {\it et al.} reported discovery of K$_{x}$Fe$_{2-y}$Se$_{2}$, a new variant of iron-based high-$T_c$ superconductor made of a combination of Fe and Se rather than Fe and As \cite{Guo}.  By intercalating K into FeSe ($T_{c}\sim9$~K \cite{Hsu, McQueen}), $T_c$ exceeds 30~K \cite{Guo}.  The crystal structure \cite{Guo} is analogous to that of BaFe$_2$As$_2$-based high $T_c$ systems \cite{Rotter, Sefat}.  Subsequent studies not only confirmed the initial discovery but also demonstrated that the Rb or Cs intercalation in lieu of K also results in high-$T_c$ superconductivity \cite{Maziopa, Mizuguchi, Wen, Lei, Wang}.   Intensive bulk and spectroscopic studies are currently underway  around the world \cite{Qian, Shermadini, Yu}. 

Very recently, ARPES measurements on  K$_{x}$Fe$_{2-y}$Se$_{2}$ revealed that, unlike iron-arsenide high $T_c$ systems, all the hole bands near the center of the first Brillouin zone are filled by electrons donated by intercalated K \cite{Qian}.  In many other iron-arsenide systems, the presence of the Fermi surface nesting between the hole bands at the zone center and additional electron bands near the zone edge, and antiferromagnetic spin fluctuations (AFSF) which may be associated with the nesting effects, have been considered an important ingredient of the high-$T_c$ mechanisms.  As such, the new   K$_{x}$Fe$_{2-y}$Se$_{2}$ system may provide a counter example to the significance of the presence of the hole bands.  

These remarkable findings have opened a new avenue to shed light on the mechanism of high $T_c$ superconductivity in iron-based systems through comparison between FeSe- and FeAs-based superconductors.  What is the key effect of the K doping that results in the elevated $T_c$?  Do the vacancies of the K sites affect the electronic properties?  Are AFSF enhanced by electron doping?  What is the nature of the superconducting state below $T_c$?  

In what follows, we will report comprehensive microscopic investigations into the structural, electronic, and superconducting properties of a single crystal sample of K$_{x}$Fe$_{2-y}$Se$_{2}$ based on $^{77}$Se NMR spectroscopy.   $^{77}$Se is an NMR active nucleus with nuclear spin $I=\frac{1}{2}$ and nuclear gyromagnetic ratio $\gamma_{n}/2\pi = 8.118$~MHz/T, and is readily observable from a relatively small crystal ($\sim 30$~mg).  Our NMR data will provide us with rich information and insight into the fascinating new model system. 

The rest of this paper is organized as follows.  In section II, we will briefly describe the experimental procedures.  Section III contains experimental results and discussions, and comparison with earlier NMR works on various iron-based superconductors.  We summarize and conclude in Section IV.

\section{\label{sec:level1}Experimental}
Single crystalline samples of nominal composition K$_{0.8}$Fe$_{2}$Se$_{2}$ were grown using the self-flux method. Powder X-ray diffraction on ground samples were taken with Cu K$_{\alpha}$ radiation ($\lambda$=1.5418 $\AA$) using a Rigaku Miniflex X-ray machine. The lattice parameters were obtained using the Rietica software \cite{Hunter}.  The elemental analysis was performed using an energy-dispersive x-ray spectroscopy (EDX) in a JEOL JSM-6500 scanning electron microscope. The result showed stoichiometry consistent with both K and Fe defficiency: K$_{0.65(3)}$Fe$_{1.41(4)}$Se$_{2.00(4)}$. We will describe the details of the crystal growth, structural refinement, and the characterization of the bulk superconducting properties elsewhere \cite{Lei, Wang}. We selected one piece of crystal weighing approximately 30~mg for our NMR measurements.  K$_{x}$Fe$_{2-y}$Se$_{2}$ (as well as FeSe) is highly sensitive to oxidizing conditions, and to avoid decomposition of the crystal, we stored and transported it in a quartz tube sealed in vacuum.  We also limited the exposure of the NMR crystal to air to several minutes during the preparation for NMR measurements.  

To confirm the high quality of the particular piece of crystal selected for NMR, we checked the superconducting transition with high frequency AC magnetic susceptibility measurements {\it in-situ} by observing the temperature dependence of the tuning frequency of our NMR circuit near 67.7~MHz.  Our NMR sample has a $T_{c}$ of 33~K in zero applied field.

All $^{77}$Se NMR measurements were conducted by applying an external magnetic field of 8.3~Tesla along the crystal c-axis or within the ab-plane.  Very narrow NMR lineshapes with the FWHM (Full Width at Half Maximum) as narrow as 4.5~kHz allowed us to measure the lineshapes by taking the Fast Fourier Transformation (FFT) of the envelope of Hahn's spin-echo.  In order to measure the spin-lattice relaxation rate $1/T_{1}$, we applied an inversion $\pi$ pulse prior to the spin-echo sequence.  We found that the recovery after the $\pi$-pulse fits nicely with the standard single exponential form, as expected for the case of nuclear spin 1/2 in a homogeneous sample.     

\section{\label{sec:level1}Results and Discussions}
\subsubsection{in-situ AC-Susceptibility Measurements}
We summarize the results of {\it in-situ} AC magnetic susceptibility measurements in Fig.\ 1.  We tuned the NMR tank circuit slightly above $T_c$ at a frequency of $f_{tune}(T_{c})\sim 67.7$~MHz, and monitored the shift of the tuning frequency $\Delta f_{tune}(T)$ as a function of temperature using a network analyzer.  The tuning frequency varies as
\begin{equation}
f_{tune}(T) \sim \frac{1}{\sqrt{LC}},
\label{1}
\end{equation}
where $L$ and $C$ represent the inductance and the capacitance of the NMR tank circuit, respectively.   When the K$_{x}$Fe$_{2-y}$Se$_{2}$  crystal undergoes a superconducting phase transition, the bulk magnetic susceptibility $\chi_{bulk}$ becomes negative because of the Meissner effects.  Accordingly, the total inductance $L=L_{o}(1+\chi_{bulk})$ also decreases below $T_c$, where $L_{o}$ is the bare inductance without the sample.  Hence $f_{tune}(T)$ increases below $T_c$ ({\it i.e.} $-\Delta f_{tune}(T) < 0$).  We note that this is a very effective technique to determine $T_c$ of the NMR sample, as demonstrated earlier for our NMR study of FeSe under pressure \cite{Imai}.  

\begin{figure}[b]
\includegraphics[width=3.2in]{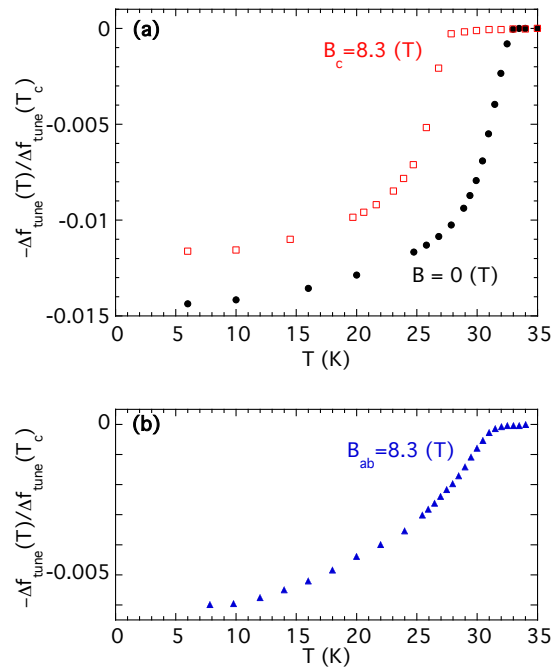}\\
\caption{\label{Fig.1} (Color Online) The change of the tuning frequency $-\Delta f_{tune}(T)$ of the NMR tank circuit induced by superconducting Meissner effects.  The data points shown, $-\Delta f_{tune}(T)/f_{tune}(T_{c})$, are normalized with the normal state value of $f_{tune}(T_{c}) \sim 67.7$~MHz .   The overall magnitude of the results in (a) and (b) should not be directly compared, because we used different coil configurations for the measurements with $B_{ab}=8.3$~Tesla.}
\end{figure}

We normalize $-\Delta f_{tune}(T)$ by dividing it with $f_{tune}(T_{c})$, and plot $-\Delta f_{tune}(T)/f_{tune}(T_{c})$ as a function of temperature in Fig.1.  The zero-field results exhibit a reasonably sharp superconducting phase transition with an onset as high as at $T_{c}=33$~K.  In the presence of an applied magnetic field of $B_{ab}=8.3$~Tesla along the ab-plane, $T_{c}$ decreases somewhat to $\sim32$~K.  In contrast, when we applied $B_{c}=8.3$~Tesla oriented along the c-axis, we observed a small but noticeable change of $f_{tune}(T)$ beginning as high as $T_{c}\sim31$~K, followed by much more dramatic change at a substantially lower temperature $\sim28$~K.  Our finding of the broad transition with a magnetic field along the c-axis may be related to a recent report that the resistive superconducting transition becomes very broad under $B_c$ due to strong two-dimensionality of the K$_{x}$Fe$_{2-y}$Se$_{2}$ system \cite{Mizuguchi}.

Because of the aforementioned sensitivity of the crystal to oxidization in air, we were forced to complete setting up our cryogenic NMR probe, NMR coil, the sample holder, and their loading into our crystat in less than ten minutes after we broke the vacuum seal.  As such, we were unable to optimize the shape of the NMR coil, and minimize its volume.  We estimate that the sample volume is as little as 10~\% or less of the total volume of our NMR coil.   Our use of such a large coil may be partially responsible for the relatively small frequency change $-\Delta f_{tune}(T)/f_{tune}(T_{c}) \sim 0.015$.  On the other hand, specific heat measurements did not reveal any anomaly at $T_c$ for a crystal selected from the same batch.  These results might suggest that the concentration of K and/or Fe is not optimized for  a highest $T_c$ or the largest superconducting volume fraction.  

\subsubsection{$^{77}$Se NMR Lineshapes}
In Fig.~2, we present representative $^{77}$Se NMR lineshapes observed at various temperatures under a magnetic field of $B_{c}=8.3$ Tesla applied along the crystal c-axis.  Except below $T_c$, the integrated intensity of the NMR lineshape  doesn't show any noticeable temperature dependence.  It is worth emphasizing that the NMR linewidth is as narrow as $\sim 4.5$~kHz in the normal state above $T_c$.  These lineshapes are by a factor of $\sim 8$ narrower than the $^{75}$As NMR central  line in the optimally Co-doped superconducting system Ba(Fe$_{1.92}$Co$_{0.08}$)$_{2}$As$_{2}$ \cite{Ning1}.  The narrowness of the $^{77}$Se lineshapes is in part due to the fact that $^{77}$Se nuclear spin is 1/2, and therefore there are no nuclear quadrupole effects.  However, such narrow $^{77}$Se NMR lineshapes also assure us that there is no macroscopic inhomogeneity in our crystal induced by, among other possibilities, a distribution in the concentration of K defects.  Given the sizable magnitude of the NMR Knight shift $^{77}K$ to be presented in Fig.\ 4, a macroscopic inhomogeneity in the K concentration would significantly broaden the $^{77}$Se lineshape though a distribution in $^{77}K$.

\begin{figure}[b]
\includegraphics[width=3.4in]{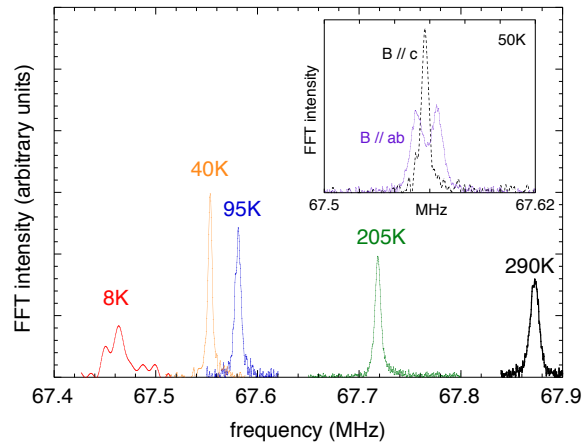}\\
\caption{\label{Fig.2} (Color Online) The representative $^{77}$Se NMR FFT lineshapes observed at various temperatures in an external magnetic field $B_{c}=8.3$~Tesla applied along the crystal c-axis.  We normalized the magnitude of each FFT trace so that the integrated intensity is equal between different temperatures.  A noisy lineshape at 8~K reflects the extremely weak signal intensity due to the superconducting shielding effects.  The systematic decrease of the peak frequency reflects the reduction of the local spin susceptibility.  Inset: a representative FFT lineshape observed at 50~K with an external magnetic field $B_{ab}=8.3$~Tesla applied along the crystal ab-plane.  For comparison, a c-axis trace observed at 50~K is also presented.}  
\end{figure}

For comparison, we also show an NMR lineshape observed at 50~K with  $B_{ab}=8.3$ Tesla applied along an arbitrary direction within the ab-plane.  The overall temperature dependence of the ab-plane lineshape was found to be very similar to the case of the c-axis lineshapes.  Unlike the case of c-axis lineshapes, however, the ab-plane lineshapes always split into two separate lines with  identical integrated intensities.   We ruled out the possibility of a mixed-phase sample based on powder x-ray diffraction measurements.  In addition, given that the two separate lines have the same integrated intensity, it is highly unlikely that the NMR line splitting is caused by the presence of two domains with different K or Fe compositions.  Furthermore, if our crystal was a mixed phase, then the c-axis NMR lineshapes would split into two as well.  We also confirmed that the separation of the two peaks ($\sim12$~kHz in $B_{ab}=8.3$~Tesla) changes in proportion to the magnitude of $B_{ab}$.  This means that the splitting is caused by paramagnetic effects of two structurally inequivalent Se sites, rather than static hyperfine magnetic fields.  Therefore we rule out the possibility that the observed line splitting is caused by a magnetic long range order. 

These considerations lead us to suggest the possibility that, locally, there are two distinct environments for Se sites under the presence of the in-plane magnetic field.  For example, suppose K defects form a superstructure and the four-fold symmetry at the Se sites is {\it locally} lowered to two-fold symmetry without breaking the overall tetragonal symmetry of the average structure.  Then the $^{77}$Se NMR line may split into two under the presence of  an in-plane magnetic field, while the c-axis lineshapes would remain single-peaked.  We remind the reader of a precedent that $^{23}$Na, $^{59}$Co and $^{17}$O NMR lines exhibits complicated line splittings in Na$_{x}$CoO$_{2}$ because of the vacancy ordering at the Na sites \cite{Alloul, NingCo, NingO}.   In fact, after the initial submission of this work, single crystal x-ray measurements showed the formation of a superstructure of K and Fe defects \cite{Zavalij}.  

\subsubsection{Estimation of the Superconducting Penetration Depth $\lambda_{ab}$ and the Carrier Density $n_{s}$ from the NMR Linewidth}
The temperature dependence of the FWHM of the NMR lineshapes, $\Delta f_{NMR}$, is summarized in Fig. 3 for the single peak observed for the c-axis measurements.  $\Delta f_{NMR}$ shows a mild increase with temperature  from  as narrow as $\Delta f_{NMR} = 4.5$~kHz near $T_c$ to 8~kHz at 290~K, the highest temperature we reached.  This increase with temperature is attributed to the increase of the NMR Knight shift $^{77}K$ in the same temperature range, as shown in Fig.\ 4.  

\begin{figure}[b]
\includegraphics[width=3.2in]{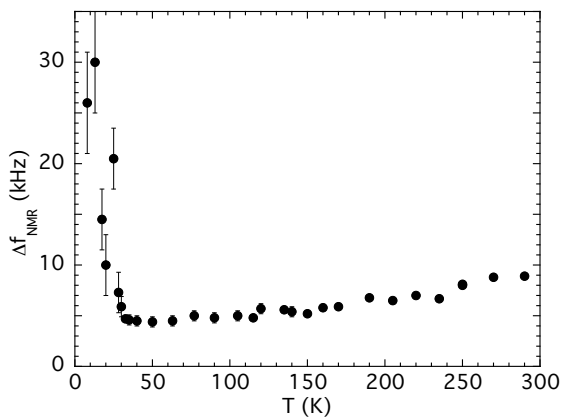}\\
\caption{\label{Fig.3} The full width at half maximum (FWHM), $\Delta f_{NMR}$, of the $^{77}$Se NMR FFT lineshapes observed in $B_{c}=8.3$~Tesla.  The dramatic enhancement below $T_c$ is caused by the Abrikosov lattice of superconducting vortices, and the enhanced width is proportional to $1/\lambda_{ab}^{2}$.}
\end{figure}

Upon entering the superconducting state, $\Delta f_{NMR}$ suddenly begins to show a dramatic increase, and reaches $\Delta f_{NMR} \sim 30$~kHz at the base temperature of our measurements, 8~K.  This is a generic feature which is expected for type-II superconductors; when the applied magnetic field penetrates our crystal along the c-axis, an Abrikosov lattice of superconducting vortices form; the magnitude of the local magnetic field varies position by position, because the magnetic field near the superconducting flux $\phi_{o} = hc/2e = 2.07 \times 10^{-7}$ Oe$\cdot$cm$^{2}$ decays with a length scale set by the in-plane {\it London penetration depth} $\lambda_{ab}$.  If the position dependent distribution of the local magnetic field $B_c$ yields a second moment $\Delta B_{c}$, the NMR linewidth will also have an enhanced second moment,  $\Delta f_{NMR}^{Vortex} = \gamma_{n} \Delta B_{c}$.  Theoretically, one can relate $\Delta B_{c}$ with $\lambda$ as \cite{Pincus, Brandt}
\begin{equation}
          \Delta B_{c} \sim  0.0609 \frac{\phi_{o}}{\lambda^{2}}.
\label{2}
\end{equation}
One can thus deduce $\lambda_{ab}$ from the NMR line broadening below $T_c$.  We note that $\mu$SR measurements of the penetration depth rely on this technique originally developed for NMR.  From the observed increase of the NMR linewidth, we estimate $\lambda_{ab} \sim 290$~nm at 8~K.  The observed value is comparable to the case of LaFeAsO$_{0.91}$F$_{0.09}$ \cite{Ahilan}.  We must note, however, that our results were obtained in a magnetic field as large as 8.3~Tesla, and application of such a high magnetic field tends to underestimate $\Delta f_{NMR}$, hence our $\lambda_{ab}$   may be somewhat overestimated.  

From the London relation, we can also estimate the density of superconducting electrons $n_{s}$ from $\lambda_{ab}$ as 
\begin{equation}
          n_{s}=\frac{m^{*}c^{2}}{4\pi e^{2} \lambda_{ab}^{2}}.
\label{3}
\end{equation}
By taking the enhancement of the effective mass $m^{*}$ as $2\sim4$ \cite{Sebastian, Ding}, we obtain $n_{s} \sim (0.7 \sim 1.2) \times 10^{+21}$~cm$^{-3}$.

\subsubsection{Uniform Spin Susceptibility as Measured by NMR Knight Shift $^{77}K$}
In Fig. 4, we summarize the temperature dependence of the $^{77}$Se NMR Knight shift $^{77}K$,
\begin{equation}
          ^{77}K = A_{hf}\chi_{spin} + K_{chem}, 
\label{4}
\end{equation}
where $A_{hf}$ is the hyperfine coupling constant between the observed $^{77}$Se nuclear spins and electron spins in their vicinity, $\chi_{spin}$ is the spin susceptibility of those electrons, and $K_{chem}$ is the temperature independent chemical shift originating from the orbital motion of electrons.  From the asymptotic value of $^{77}K$ observed well below $T_c$, we estimate $K_{chem} \sim 0.11$~\%.  Since Fe-based superconductors are multi-orbital systems, in principle,  different orbitals may have separate contributions to both $A_{hf}$ and  $\chi_{spin}$.  In what follows, we ignore such potential complications for the sake of  simplicity.  We need to bear in mind, however, that $^{77}K$ may be a weighted average of the spin susceptibility of different orbitals.

\begin{figure}[b]
\includegraphics[width=3.4in]{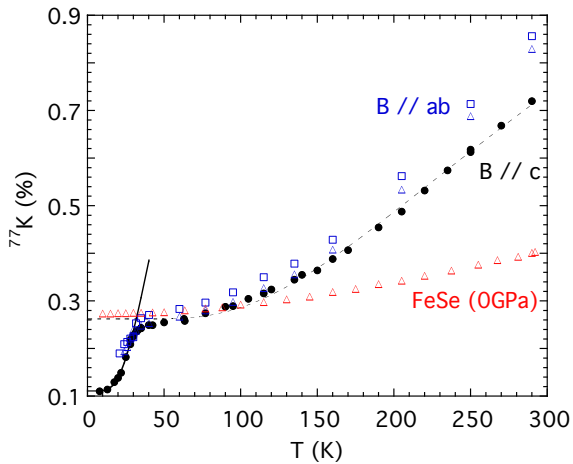}\\
\caption{\label{Fig4} (Color Online) The temperature dependence of the uniform ${\bf q}={\bf 0}$ component of the static magnetic spin susceptibility $\chi_{spin}$ as measured by $^{77}$Se NMR Knight shift,  $^{77}K$.  Filled and open symbols represent the results with magnetic field 8.3 T applied along the crystal c-axis ( $^{77}K_c$) and ab-plane ($^{77}K_{ab}$), respectively.  Notice a sharp drop below $T_c$ for both orientations, and the change of the curvature in the normal state near 200~K.   The solid curve represents a fit with an activation form, $^{77}K_{spin}  \sim exp (-\Delta_{sc}/k_{B}T)$, with the superconducting energy gap $\Delta_{sc}/k_{B} = 92$~K.   The dashed curve represents a phenomenological fit with a pseudo-gap, $^{77}K_{spin} \sim exp (-\Delta/k_{B}T)$, with $\Delta/k_{B} = 435$~K.  For comparison, we also present the normal state results above $T_{c}=9$~K of FeSe \cite{Imai}. }
\end{figure}

Fig.\ 4 presents both $^{77}K$ measured with a magnetic field applied along the c-axis ( $^{77}K_c$) and ab-plane ($^{77}K_{ab}$).  $^{77}K_c$ and $^{77}K_{ab}$ show qualitatively the same temperature dependence in the entire temperature range.  Plotted in Fig.\ 5 is $^{77}K_{ab}$ vs. $^{77}K_c$ by choosing temperature as the implicit parameter, and the fit with a straight line is good.  This linearity implies that the spin susceptibility along the c-axis, $\chi^{c}_{spin}$, indeed shows identical behavior to $\chi^{ab}_{spin}$ along the ab-plane.  Within experimental uncertainties, the slope of the fits are identical for the two separate peaks of the ab-plane lineshape,  $1.21\pm0.02$.  If we {\it assume}  that the hyperfine coupling is isotropic and does not depend on direction ({\it i.e.} $A^{c}_{hf}=A^{ab}_{hf}$), the anisotropy of the spin susceptibility may be estimated as $\chi^{ab}_{spin}/\chi^{c}_{spin}=1.21\pm0.02$.   

\begin{figure}[b]
\includegraphics[width=3.2in]{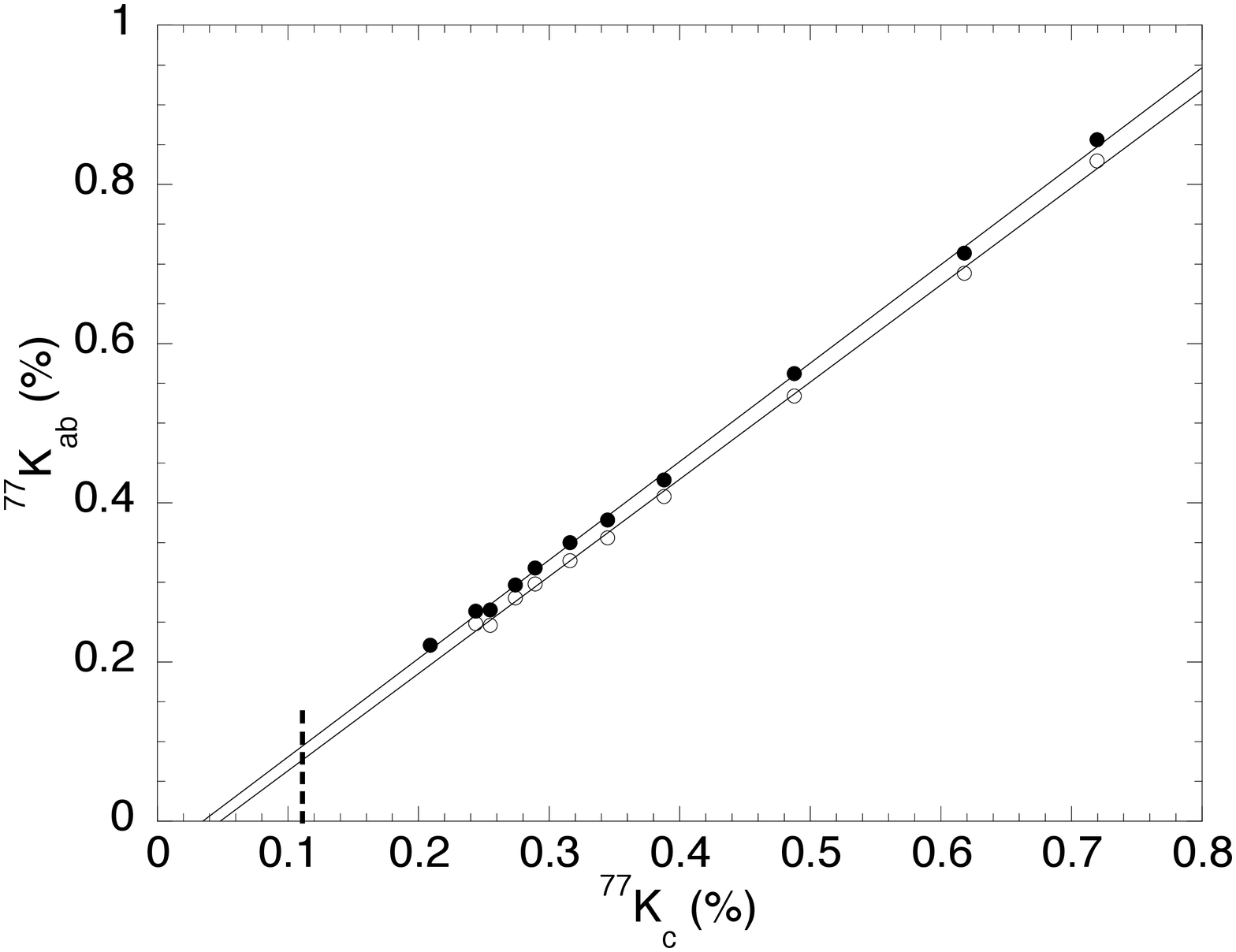}\\
\caption{\label{Fig5} $^{77}K_{ab}$ plotted as a function of $^{77}K_{c}$ by choosing temperature as the implicit parameter.  We find a good linear fit with a slope of  $1.22\pm0.01$, implying an identical temperature dependence of $\chi^{ab}_{spin}$ and $\chi^{c}_{spin}$, but with a possibility of a sizable anisotropy in their magnitudes.  Vertical dashed line represents a provisional estimation of the chemical shift $^{77}K_{chem,c}=0.11 \pm 0.02$~\% along the c-axis.}
\end{figure}

Let's turn our attention to the temperature dependence of $\chi_{spin}$ as reflected by $^{77}K$ measurements.  In the normal state above $T_c$, $\chi_{spin}$ decreases monotonically with temperature.  Similar behavior was previously reported for many iron-based superconductors, including LaFeAsO$_{1-x}$F$_{x}$ \cite{Ahilan, Nakai}, Ba(Fe$_{1-x}$Co$_{x}$)$_2$As$_2$ \cite{USTC, Ning1, Ning2, Ning3}, and FeSe \cite{Imai}.  In addition, close inspection of Fig.\ 4 reveals that the curvature of $^{77}K$ changes from positive to negative around 200~K, and hence $\chi_{spin}$ may be gradually saturating at higher temperatures.   We observe an analogous tendency in the temperature dependence of $1/T_{1}T$ as shown below.  Overall, the temperature dependence of $\chi_{spin}$ in the present case is similar to that of LaFeAsO$_{1-x}$F$_{x}$ \cite{Ahilan, Nakai}.

The mechanism of the suppression of  $\chi_{spin}$ above $T_c$ has been highly controversial.     In view of the proximity between the superconducting and SDW ordered states in the phase diagram of iron-based systems, it is natural to speculate that antiferromagnetic short range order may be growing toward $T_c$ and suppressing $\chi_{spin}$ \cite{USTC}.  Early observation of the enhancement of short range antiferromagnetic spin correlations near $T_c$ by the measurements of $1/T_{1}T$ in the optimally doped superconducting phase also seem to support such a scenario \cite{Ning1, Ning2}.  On the other hand, even the heavily over-doped, non-superconducting, non-magnetic metallic phase Ba(Fe$_{1-x}$Co$_{x}$)$_2$As$_2$ with Co concentrations as high as 20 \% or greater also turned out to show analogous behavior \cite{Ni, Ning3}.  

Since the SDW phase is located far away from the over-doped regime in the electronic phase diagram, the latter observation may point toward other scenarios, such as the change of the effective density of states of low energy spin excitations.  In canonical Fermi liquid systems, if the chemical potential $\mu$ is located on the lower energy side of a sharp peak of the density of states,  $\chi_{spin}$ could decrease at lower temperatures when $N^{*}(\mu)$ becomes smaller.  In this scenario, our results in Fig.\ 4 imply that there is a large, sharp peak of $N^{*}(E)$ somewhat above the Fermi energy $E_F$.  Recalling the mild saturating tendency of  $^{77}K$ above 200~K ($\sim 18$~meV), such a peak may be located fairly close to $E_F$.  However, the recent ARPES measurements \cite{Qian} don't seem to indicate the presence of such a peak.  

Alternatively, if one believes that the temperature dependence of $\chi_{spin}$ is dominated by frustration effects between nearest and next-nearest neighbor Fe-Fe exchange interactions, it is conceivable that more exotic {\it pseudo-gap}-like behavior emerges upon cooling the system.  We may deduce such a phenomenological pseudo-gap by fitting the spin contribution $^{77}K_{spin}$ to an activation form, $^{77}K_{spin} = ^{77}K -^{77}K_{chem} \sim exp (-\Delta/k_{B}T)$, as shown by a solid curve in Fig.\ 4.  Notice that our phenomenological fit successfully reproduces the saturating tendency of $^{77}K_{c}$ near 290~K as well.  The best fit of our data from $T_c$ to 290~K yields $\Delta/k_{B}=435$~K.  This value is comparable to the previous estimate of  $\Delta/k_{B}=450 \sim 711$~K in Ba(Fe$_{1-x}$Co$_{x}$)$_2$As$_2$ \cite{Ning2, Ning3}, and much greater than $\Delta/k_{B}=140 \sim 172$~K in LaFeAsO$_{1-x}$F$_{x}$ \cite{Nakai, ImaiTokyo}.

Below $T_c$, $^{77}K$ suddenly begins to decrease from $^{77}K_{c} \sim 0.25$~\%, then levels off at $K_{chem}=0.11\pm0.01$~\%.   We confirmed that $^{77}K_{ab}$ also decreases immediately below $T_c$ for both of the peaks.  The presence of the line splitting makes accurate measurements near the base temperature more difficult  for $^{77}K_{ab}$, because two peaks become almost indistinguishable due to NMR line broadening caused by the Abrikosov lattice.  These observations in the superconducting state below $T_c$ are consistent with  the suppression of both  $\chi^{ab}_{spin}$ and $\chi^{c}_{spin}$ expected for the singlet pairing of Cooper pairs.   

In order to estimate the magnitude and symmetry of the superconducting energy gap from the temperature dependence of  $^{77}K_{c}$ below $T_c$, one needs to estimate the diamagnetic effects, a formidable task in many superconductors \cite{Barret}.  Our application of as high as 8.3~Tesla of magnetic field may also skew the temperature dependence of $^{77}K_{c}$ to some extent by destroying the Cooper pairs.  We would need a much larger NMR signal intensity to conduct NMR measurements below $T_c$ at lower magnetic fields.  By ignoring all these potential complications, and assuming the same magnitude of isotropic energy gaps for all pieces of the Fermi surface sheets, we attempted to fit the low temperature behavior of $^{77}K_{c}$ with an activation law, $^{77}K_{spin} \sim exp(- \Delta_{sc}/k_{B}T)$.  A provisional estimate of the superconducting energy gap is $\Delta_{sc}/k_{B} \sim 92$~K, hence we obtain a much larger ratio $\Delta_{sc}/k_{B}T_{c} \sim 2.8$ than the BCS value of 1.76.   The comparatively large gap should be taken with a grain of salt in view of the level of uncertainty in our estimation of $K_{chem} \sim 0.11$~\%. 

\subsubsection{Low Energy Spin Excitations as Measured by NMR Spin-Lattice Relaxation Rate $1/T_{1}$}
Fig.\ 6 presents the temperature dependence of the $^{77}$Se nuclear spin-lattice relaxation rate $^{77}(1/T_{1})$ measured in $B_{c}= 8.3$~T applied along the c-axis.  Preliminary measurements in $B_{ab}= 8.3$~T applied along the ab-plane showed nearly identical behavior at least up to 160~K.  $^{77}(1/T_{1})$ dives below $T_c$, without exhibiting a Hebel-Slichter coherence peak expected for conventional isotropic BCS s-wave superconductors.   Due to the extremely poor signal-to-noise ratio well below $T_c$ (see the lineshape at 8~K in Fig.~2), we didn't attempt to follow the temperature dependence of $^{77}(1/T_{1})$ to much below $T_c$.  In any event, vortex motions and additional relaxation processes at the normal vortex cores generally make the interpretation of the high-field NMR measurements of $1/T_{1}$ a highly complicated affair well below $T_c$ \cite{McLaughlin}.
\begin{figure}[b]
\includegraphics[width=2.8in]{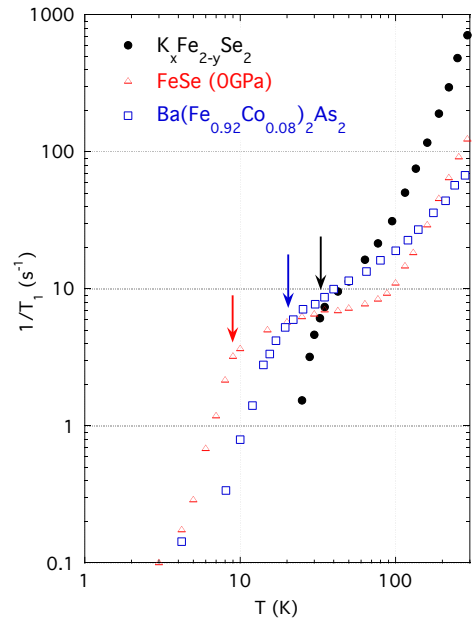}\\
\caption{\label{Fig6} (Color Online) The temperature dependence of $1/T_{1}$ observed at $^{77}$Se sites of K$_{x}$Fe$_{2-y}$Se$_2$ is compared with that observed for $^{77}$Se sites of FeSe ($T_{c}=9$~K in 0GPa) \cite{Imai} and $^{75}$As sites of Ba(Fe$_{0.92}$Co$_{0.08}$)$_{2}$As$_{2}$ ($T_{c}=25$~K) \cite{Ning1}.  Arrows mark the $T_{c}$ for each sample.}
\end{figure}
Analogous behavior of $1/T_{1}$ immediately below $T_c$ was previously reported for many other iron-based high $T_c$ superconductors \cite{Nakai, Matano, Grafe, Yashima, Ning1, Tou, Imai}.  We reproduce some of our earlier results in Fig.\ 5 for comparison.  

Mazin {\it et al.} pointed out that the lack of the Hebel-Slichter coherence peak does not necessarily mean that these systems are unconventional superconductors with line or point nodes in the superconducting energy gaps \cite{Mazin}.  That is, even if the energy gaps are fully open, alteration of the sign of the superconducting energy gaps between different pieces of the Fermi surface could lead to annihilation of the coherence peak in the so-called s$_{\pm}$ states.  They considered several s$_{\pm}$ states in the context of LaFeAsO$_{1-x}$F$_{x}$, which has both electron Fermi surfaces near the zone edge and the hole Fermi surfaces at the zone center.  In the present case, recent ARPES measurements suggest that the hole Fermi surface at the zone center is completely filled by doped electrons, and only electron Fermi surfaces exist at zone edges \cite{Qian}.  Within the s$_{\pm}$ scenario, our observation of the lack of the coherence peak therefore requires sign reversal of the energy gap between the adjacent pieces of the electron Fermi surface at zone edges.  This requires a two-fold symmetry in the crystal structure, and vacancy ordering may be playing a key role in lowering the tetragonal symmetry, as discussed above. 

Presented in Fig. 7 is the temperature dependence of $^{77}(1/T_{1}T)$, {\it i.e.} $^{77}(1/T_{1})$ divided by temperature $T$.  Quite generally, $^{77}(1/T_{1}T)$ measures the ${\bf q}$-summation of the imaginary part of the dynamical electron-spin susceptibility weighted by the hyperfine form-factor $|A_{hf}({\bf q})|^{2}$,
\begin{equation}
	1/T_{1}T \propto \sum_{{\bf q}\in B.Z.} {|A_{hf}({\bf q})|^{2} \frac {\chi " ({\bf q}, f_{NMR})}  {f_{NMR}}},
\label{5}
\end{equation}
where $f_{NMR}$ is the NMR frequency.  In other words, $^{77}(1/T_{1}T)$ probes various ${\bf q}$-modes of the low frequency spin fluctuations at the low energy scale ($hf_{NMR}$) set by the NMR frequency.  Our results show that $^{77}(1/T_{1}T)$ decreases with temperature from 290~K down to $T_c$.  The observed temperature dependence is qualitatively similar to that of $^{77}K$, including the mild saturating tendency near 290~K.  Since  $^{77}(1/T_{1}T)$ measures all wave-vector ${\bf q}$ modes of spin fluctuations while $^{77}K$ picks up the uniform ${\bf q}={\bf 0}$  mode exclusively, there is no {\it a priori} reason to expect such analogous behaviors.  In fact, in many iron-based high $T_c$ superconductors,  $1/T_{1}T$ and $K$ exhibit qualitatively different temperature dependences near $T_c$.  

For comparison, we present $^{77}(1/T_{1}T)$ observed for FeSe \cite{Imai} in Fig7(a), and $^{75}(1/T_{1}T)$ measured at $^{75}$As sites in two representative compositions of Ba(Fe$_{1-x}$Co$_{x}$)$_{2}$As$_{2}$ \cite{Ning3} in Fig.7(b).  $1/T_{1}T$ is enhanced toward $T_c$ in both FeSe ($T_{c}=9$~K) and the optimally doped Ba(Fe$_{0.92}$Co$_{0.08}$)$_{2}$As$_{2}$ ($T_{c}=25$~K), even though the  uniform ${\bf q}={\bf 0}$  mode of spin susceptibility stays nearly flat near $T_c$ in these systems.  This enhancement of $1/T_{1}T$ near $T_c$ provides evidence for enhancement of the low frequency components of the antiferromagnetic spin fluctuations (AFSF) toward $T_c$ at some finite wave vectors in FeSe and Ba(Fe$_{0.92}$Co$_{0.08}$)$_{2}$As$_{2}$.  In contrast, our results indicate that this is not the case in K$_{x}$Fe$_{2-y}$Se$_{2}$.  Instead, our new results are qualitatively similar to overdoped, non-supercondcting Ba(Fe$_{0.86}$Co$_{0.14}$)$_{2}$As$_{2}$, as shown in Fig.\ 7(b). The absence of AFSF in the latter was interpreted as the consequence of the lack of the nested Fermi surfaces between the hole and electron bands, because the overdoped electrons fill up the hole bands at the zone center \cite{Ning3}.   Since ARPES data suggest that K$_{x}$Fe$_{2-y}$Se$_{2}$  does not possess a hole Fermi surface near the B.Z. center, the absence of low frequency AFSF itself may be consistent with such an interpretation.  However, our results inevitably raise the question of whether the low frequency components of AFSF are relevant to the superconducting mechanism.  Needless to say, our NMR results do not exclude the possibility that AFSF is being enhanced at higher energies than $hf_{NMR}$, and such modes may be related to the superconducting mechanism.  We also caution that the signature of a mild enhancement of AFSF toward $T_c$ for a finite ${\bf q}$ mode could be easily masked, if the overwhelmingly large background contributions from all other ${\bf q}$ modes keep decreasing toward $T_{c}$.

\begin{figure}
\includegraphics[width=3.2in]{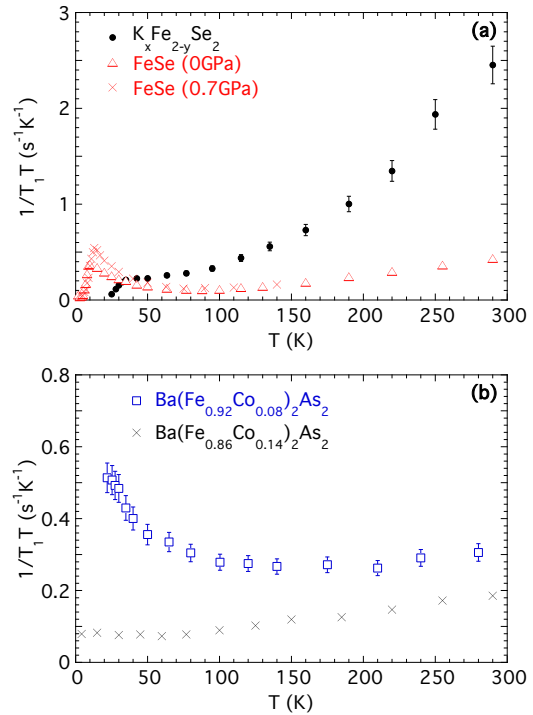}\\
\caption{\label{Fig7} (Color Online) (a) The temperature dependence of $^{77}(1/T_{1}T)$ observed for K$_{x}$Fe$_{2-y}$Se$_{2}$ is compared with the results for FeSe in 0GPa ($T_{c}=9$~K) and 0.7GPa ($T_{c}=14$~K in 0.7GPa) \cite{Imai}.  (b) Representative results of $^{75}(1/T_{1}T)$ observed at $^{75}$As sites of Ba(Fe$_{1-x}$Co$_{x}$)$_{2}$As$_{2}$ with $x=0.08$ ($T_{c}=25$~K) and $x=0.14$ ($T_{c}\sim 0$~K) with an external magnetic field applied along the ab-plane \cite{Ning3}.  Notice the  qualitative similarity between K$_{x}$Fe$_{2-y}$Se$_{2}$ and the overdoped Ba(Fe$_{0.86}$Co$_{0.14}$)$_{2}$As$_{2}$; both of these materials lack the hole bands near the zone center.}
\end{figure}

\begin{figure}
\includegraphics[width=3.7in]{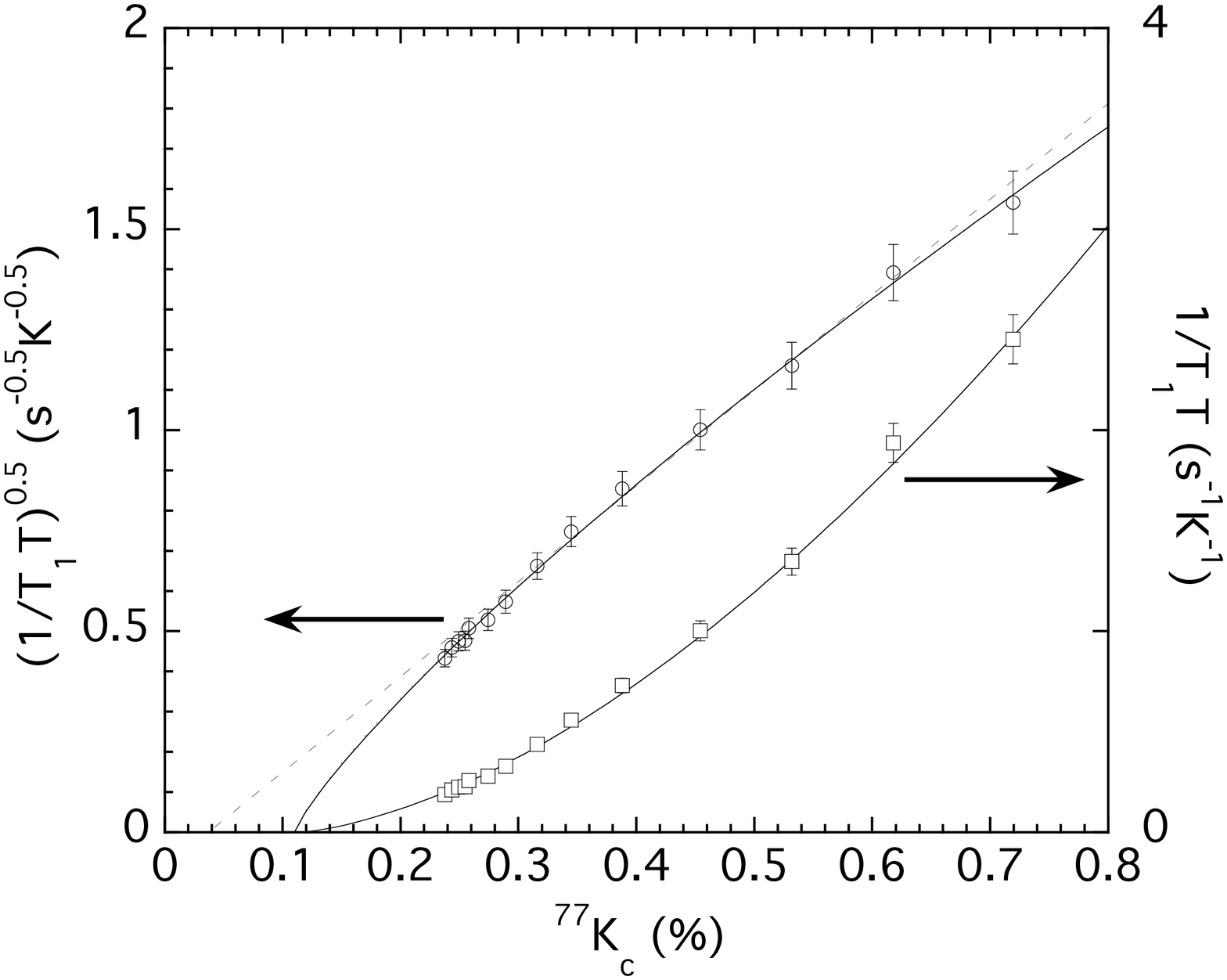}\\
\caption{\label{Fig8} The normal state data of $^{77}(1/T_{1}T)$ and its square-root, $^{77}(1/T_{1}T)^{0.5}$, are compared with $^{77}K_c$ by choosing temperature as the implicit parameter.  The dashed line represents a possible Korringa relation.  Solid curves represent free parameter fits, with $^{77}(1/T_{1}T) = (^{77}K_{c} - K_{chem})^{n}$, which yields $n \sim 1.6$ instead of the Fermi liquid value of $n \sim 2$.}
\end{figure}

Finally, let us re-examine more global features of $^{77}(1/T_{1}T)$ and $^{77}K$ above $T_c$ up to 290~K in light of the qualitative similarity in their temperature dependences.  A possible explanation for the similar behavior is that $^{77}(1/T_{1}T)$ as well as  $^{77}K$ is dominated by the effective density of states $N^{*}(\mu)$ at the chemical potential of a Fermi liquid.  Since $^{77}(1/T_{1}T) \propto [N^{*}(\mu)]^{2}$ and $^{77}K_{spin}\propto N^{*}(\mu)$, one would expect a {\it Korringa relation}, $^{77}(1/T_{1}T)^{0.5} \propto ^{77}K_{spin}$, for a Fermi liquid.  To test such a scenario, we plot $^{77}(1/T_{1}T)^{0.5}$ as a function of $^{77}K$ in Fig.~8 by choosing temperature as the implicit parameter.  It is not impossible to draw a straight line through all the data points above $T_c$, as shown by a dashed line.  However, our high precision data deviate consistently from the straight line with a negative curvature.  In addition, an extrapolation of the straight line yields the intercept of the horizontal axis at  $^{77}K\sim 0.04$~\%.  That is, the Korringa relation for a Fermi liquid would require $K_{chem} \sim 0.04$~\%, which seems low compared with our experimental results near the base temperature,  $^{77}K_{c} = 0.11\pm 0.01$~\%. An alternative scaling form to account for the similarity in the temperature dependence of $^{77}(1/T_{1}T)$ and $^{77}K$ would be $^{77}(1/T_{1}T) \propto ^{77}K_{spin}$ for nearly ferromagnetic Fermi liquid, but our plot in Fig.\ 8 shows that this relation fails completely.  This is not surprising; after all, strong ferromagnetic correlation would result in enhancement of $^{77}K$ with decreasing temperature.  Instead, if we float two parameters, the power $n$ as well as $K_{chem}$ in the dynamical scaling form of $^{77}(1/T_{1}T) = (^{77}K_{spin, c})^{n} =(^{77}K_{c} - K_{chem})^{n}$, we obtain $n \sim 1.6$ and $K_{chem}\sim 0.12$~\% rather than the Fermi liquid value of $n=2$.  The solid curves in Fig.\ 8 show the scaling fits.

\section{\label{sec:level1}Summary and Conclusions}
In this paper, we have reported a thorough and detailed investigation of the structural, electronic, and superconducting properties of the newly discovered high $T_c$ superconductor K$_{x}$Fe$_{2-y}$Se$_{2}$ using $^{77}$Se NMR technuiques.  While the details of the spatial arrangements of the K and Fe defects are still unknown, our NMR lineshapes suggest that tetragonal four-fold symmetry is broken down at Se sites -- at least locally -- and two-fold symmetry in the local crystal structure may be induced by the formation of a defect superstructure.  In addition, it is worth noting that we found no evidence for coexistence of magnetically ordered phase in our NMR lineshapes, because NMR lines are very sharp and exhibit paramagnetic behavior.

Electronic properties in the normal state above $T_c$ share many common traits with the low $T_c$ superconductor FeSe as well as with iron-arsenide high $T_c$ superconductors.  For example, the uniform spin susceptibility decreases with temperature toward $T_c$, as observed for all other iron-arsenide high $T_c$ systems as well as for FeSe.  In view of the closeness of the power $n\sim 1.6$ in the dynamical scaling analysis of $^{77}(1/T_{1}T)$ vs. $^{77}K_{spin}$ to the Fermi liquid value of $n=2$, our results in Fig.\ 7 likely imply a suppression of spin excitations in a broad range of wave vector ${\bf q}$'s, with a possible pseudo-gap-like excitation gap $\Delta/k_{B} \sim 435$~K.  These NMR results are qualitatively similar to the overdoped non-superconducting system Ba(Fe$_{0.84}$Co$_{0.16}$)$_{2}$As$_{2}$, and we do not find evidence for the enhancement of AFSF near $T_c$.   In this context, it may be important to recall that the presence of AFSF in the low $T_c$ superconductor FeSe was not easy to detect, because defects of  only a few percent could wipe out the signature of AFSF while reducing the volume fraction of superconductivity \cite{McQueen, Imai, Tou}.    Moreover, if one reduces the Co concentration of Ba(Fe$_{0.84}$Co$_{0.16}$)$_{2}$As$_{2}$ by just a few percent, a robust signature of AFSF emerges with progressively larger volume fractions of superconductivity as one approaches the optimal doping concentration $x \sim 0.07$ \cite{Ning3}.  Given that the defect concentrations of K and Fe and their spatial coordination have not been fully controlled, and that the specific heat anomaly has not been observed at $T_c$ in our sample, we need to interpret the implications of our NMR results with caution.  

Turning our attention to superconductivity, we extracted the London penetration depth $\lambda_{ab} \sim 290$~nm and the carrier concentration $n_{s} \sim 1 \times 10^{+21}$ cm$^{-3}$ based on the NMR line broadening below $T_{c}$ induced by the Abrikosov lattice.   We also found that the NMR Knight shift decreases below $T_c$ for both magnetic field applied along the crystal c-axis and the ab-plane.  In addition, we showed that there is no Hebel-Slichter coherence peak in the temperature dependence of $1/T_1$ just below $T_{c}$.  These results are consistent with singlet pairing of Cooper pairs with the s$_{\pm}$ symmetry.  However, lack of accurate data well below $T_c$ prevents us from ruling out the possibility of anisotropic pairing. 

During the course of the present study, Yu {\it et al.} posted a preprint, and reported their NMR investigation of  KFe$_{2-y}$Se$_{2}$ in a limited temperature range above $T_{c}$ \cite{Yu}.  Their NMR data in the normal state agree reasonably well with our results in the temperature range overlapped.  Subsequently, during the final stage of the present work, two preprints appeared.  First, Ma, Yu {\it et al.}  posted an additional preprint, and reported an extension of their work to below $T_{c}$, with primary emphasis on $1/T_{1}$ measured in a high magnetic field \cite{Ma}. Kotegawa and co-workers also extended the earlier measurements to a broader temperature range \cite{Kotegawa}. 

\begin{acknowledgments}
Work at McMaster is supported by NSERC and CIfAR.  Work at Brookhaven is supported by the U.S. DOE under Contract No. DE-AC02-98CH10886 and in part by the Center for Emergent Superconductivity, an Energy Frontier Research Center funded by the U.S. DOE, Office for Basic Energy Science
\end{acknowledgments}


%

\end{document}